\documentclass[12pt, eqno]{article}
\usepackage[dvips]{color}
\usepackage{epsfig}
\usepackage{amsmath}
\usepackage{graphicx}
\begin{document}
\title{\bf The effect of higher derivative correction on $\eta /s$ and conductivities in STU model}
\author{{J. Sadeghi$^{a,b,}$\thanks{Email:
pouriya@ipm.ir}\hspace{1mm},
B. Pourhassan$^{a,}$\thanks{Email: b.pourhassan@umz.ac.ir}\hspace{1mm}
and A. R. Amani$^{c,}$\thanks{Email: a.r.amani@iauamol.ac.ir}}\\
$^a$ {\small {\em  Sciences Faculty, Department of Physics, Mazandaran University,}}\\
{\small {\em P .O .Box 47416-95447, Babolsar, Iran}}\\
$^b$ {\small {\em  Institute for Studies in Theoretical Physics and
Mathematics (IPM),}}\\
{\small {\em P.O.Box 19395-5531, Tehran, Iran}}\\
$^c$ {\small {\em  Islamic Azad University, Ayatollah Amoli Branch, Department of Physics, Amol, Iran}}\\
{\small {\em P.O.Box 678, Amol, Iran}} } \maketitle
\begin{abstract}
\noindent In this paper we study the ratio of shear viscosity to
entropy, electrical and thermal conductivities for the R-charged
black hole in STU model. We generalize previous works to the case of
a black hole with three different charges. Actually we use diffusion constant to obtain ratio of shear viscosity to entropy.
By applying the thermodynamical stability we recover previous results.  Also we investigate the effect of higher derivative corrections.
\end{abstract}
\newpage
\tableofcontents
\newpage
\section{Introduction}
As we know the famous example of AdS/CFT correspondence [1, 2, 3] is the relation between type IIB string theory in the $AdS_{5}\times S^{5}$ space and a
$\mathcal{N}=4$ super Yang-Mills gauge theory on the 4-dimensional boundary of $AdS_{5}$ space. This duality has been extended to a different cases. For
example it can describe confining gauge theories such as QCD. In order to have a more realistic description of the strongly coupled quark-gluon plasma
(QGP) we have to apply such duality. On the other hand, one of the remarkable link between black hole thermodynamics and field theory arise by the
holography. We note here for the providing complete information about the field theory. The AdS/CFT help us to provide dual descriptions of any desired
process in the gauge theory. We do not expect to find simple description to obtain full information of the gauge theory. In thermodynamic point of view it
is natural to expect that the long-distance fluctuations in the theory we have hydrodynamic description. In the hydrodynamics we have some important
quantities such as shear viscosity, which plays important role in physics of early universe. Such studies are important to understand the physics of the
early universe. The most important problems of QGP are the shear viscosity, drag force and jet-quenching parameter. It is found that the ratio of shear
viscosity $\eta$ to the entropy density $s$ had a universal value, i.e., $\eta/s=1/4\pi$ [4-19]. However for the general case of the coupled system it can
be shown that $\eta/s\geq1/4\pi$. Previous computation of the shear viscosity usually based on the Kubo formula . In this paper, we use diffusion constant
[4] to obtain the ratio of shear viscosity to entropy density for the three-charged black hole in the STU model with higher derivative correction. The STU
model admits a chemical potential for the $U(1)^{3}$ symmetry and this makes it more interesting. Already the shear viscosity in the STU background
computed [9,10] and higher derivative effects of the five-dimensional gauged supergravity [20] applied on the ratio of shear viscosity to entropy [21]. The
STU model is an example of $D=5$, $\mathcal{N}=2$ gauged supergravity theory [22] which is dual to the $\mathcal{N}=4$ SYM theory with finite chemical
potential. The ${\mathcal{N}}=2$ supergravity theory in five dimensions can be obtained by compactification of the eleven dimensional supergravity in a
three-fold Calabi-Yau [23]. The $D=5$, $\mathcal{N}=2$ gauged supergravity theory is a natural way to explore gauge/gravity duality, and three-charge
non-extremal black holes are important thermal background for this correspondence. For these reasons we already calculated the quantities of the drag force
and jet-quenching parameter in
the STU background [24-27].\\
Another important property of the QGP is called the jet-quenching parameter, so the knowledge about this parameter increases our understanding about the
QGP. In that case the jet-quenching parameter is obtained by calculating the expectation value of a closed light-like Wilson loop and using the dipole
approximation [28]. In order to calculate this parameter in QCD one needs to use perturbation theory. But by using AdS/CFT correspondence the jet-quenching
parameter calculated in the non-perturbative quantum field theory. This calculations performed in the $\mathcal{N}=4$ SYM thermal plasma [29-35]. In the
Ref. [27] we calculated the jet-quenching parameter in the STU background include a black hole with three charges. We should note that our paper is
extension of the Refs. [10] and [21] because we are going to consider the STU model with three different charges.\\
There are also interesting hydrodynamical quantity such as thermal
and electrical conductivity which can be calculated from
gauge/gravity duality. In the recent work [36] the thermal and
electrical conductivity calculated in the presence of non-zero
chemical potential and found that conductivities for gauge theories
dual to R-charged black hole in $d=4$ behaves in a universal manner.
In the Ref. [36] R-charged black holes in arbitrary dimension
considered and electrical conductivity computed. We use results of
the Ref. [36] to write an expression for electrical conductivity
as a hydrodynamical property of the QGP.\\
This paper organized as the following. In section 2 we give brief review of the STU model. In section 3 we calculate the ratio of shear viscosity to
entropy density by using the diffusion constant. In section 4 we obtain the effect of higher derivative corrections on the ratio of shear viscosity to
entropy density. Then in section 5 we extract thermal and electrical conductivities, and discuss the effect of higher derivative correction. In section 6
we calculate the shear viscosity to entropy ratio by using density of physical charge. Finally in section 7 we summarize our results.
\section{STU Model}
The STU model is the special form of the ${\mathcal{N}}=2$
supergravity in different dimensions, and generally has 8-charge
non-extremal black hole (4 electric and 4 magnetic). However, there
are many situations for the charge configurations such as
four-charge and three-charge black holes. There is great difference
between three-charge and four-charge black hole. For example if
there are only 3 charges in 4 dimensions, then the entropy vanishes
(except in the non-BPS case), so one really needs four charges to
get a regular black hole. In 5 dimensions the situation is different
and actually much simpler, there is no distinction between BPS and
non-BPS branch, in this dimension the three-charge configurations
are the most interesting ones [37]. Therefore, we begin with the
three-charge non-extremal black hole solution in ${\mathcal{N}}=2$
gauged supergravity which is called STU model and described by the
following solution [38],
\begin{equation}\label{1}
ds^{2}=-(H_{1}H_{2}H_{3})^{-\frac{2}{3}} f dt^{2} +(H_{1} H_{2}
H_{3})^{\frac{1}{3}} \Big(\frac{dr^{2}}{f}+r^{2}
d\Omega_{3,k}^2\Big),
\end{equation}
where,
\begin{eqnarray}\label{2}
f&=&k-\frac{\mu}{r^{2}}+\lambda^{2}r^{2}H_{1} H_{2}
H_{3},\nonumber\\
H_{i}&=&1+\frac{q_{i}}{r^{2}}, \hspace{10mm} i=1, 2, 3,
\end{eqnarray}
and,
\begin{equation}\label{3}
d\Omega_{3,k}^2=dx^2+dy^2+dz^2,
\end{equation}
where $k = +1, -1, 0$ are $S^{3}$, pseudo-sphere and flat spaces,
respectively. $\lambda\equiv\frac{1}{R}$ is the inverse curvature of
$AdS_{5}$ and related to the cosmological constant via
$\Lambda=-6\lambda^{2}$. In STU model there are three real scalar
fields as $X^{i}=\frac{{(H_{1} H_{2} H_{3})}^{\frac{1}{3}}}{H_{i}}$
which satisfy the following condition, $\prod_{i=1}^{3}X^{i}=1$. In
another word, if we set $X^{1}=S$, $X^{2}=T$ and $X^{3}=U$ then
there is $STU=1$ condition.\\
The Hawking temperature of the solution (1) is given by the
following relation,
\begin{equation}\label{4}
T=\frac{2+\frac{q_1+q_2+q_3}{r^2_h}-\frac{q_1\,q_2\,q_3}{r^6_h}}
{2\,\prod^3_{i=1}(1+\frac{q_i}{r_h^2})^{\frac{1}{2}}}\frac{r_h}{\pi}\lambda^2,
\end{equation}
also the chemical potential is given by,
\begin{equation}\label{5}
\phi_{i}^{2}=q_{i}\,\lambda^2\,(1+\frac{q_{i}}{r_{h}^{2}})\prod_{j\neq
i}(r_{h}^{2}+q_{j}).
\end{equation}
It is clear that at $q_{i}=0$ limit the chemical potential vanishes
and the Hawking temperature reduces to one in the $N=4$ SYM theory,
i.e. $T=\frac{r_h}{\pi}\lambda^2$.
\section{Ratio of shear viscosity to entropy}
In this section we are going to study universality of the ratio of shear viscosity to
entropy density, $\eta/s$. There are several ways to compute the
shear viscosity such as the Kubo formula [39, 40]. In this paper we use diffusion
constant approach to extract the ratio of shear viscosity to entropy
density. In that case for a given metric, $ds^2=g_{tt}dt^2+g_{rr}dr^2+g_{xx}d\vec{x}^2$,
the diffusion constant becomes [4],
\begin{equation}\label{6}
D=\frac{\sqrt{-g(r_{h})}}{\sqrt{-g_{tt}(r_{h})g_{rr}(r_h)}}\,
\int^{\infty}_{r_h}dr\frac{-g_{tt}\,g_{rr}}{g_{xx}\sqrt{-g}}.
\end{equation}
We should note that the equation (6) works only for the case of flat space where $k=0$. Then, in order to investigate universality of the ratio of shear
viscosity to entropy density we use $\eta/s=TD$. By using the line element (1) one can obtain the following expansion,
\begin{equation}\label{7}
D=\prod_{i}(1+\frac{q_i}{r_h^2})^{\frac{1}{6}}
\left[\frac{1}{2r_{h}}-{\frac{\sum_{i} q_{i}}{{6\,
r_{{h}}}^{3}}}+{\frac {5\sum_{i} q_{i}^{2}+ 4\sum_{i\neq j}
q_{{i}}q_{{j}}}{{54r_{{h}}}^{5}}} - \cdots\right],
\end{equation}
also by using the Hawking temperature (4) we can obtain the
following expression for the ratio of shear viscosity to entropy
density,
\begin{equation}\label{8}
\frac{\eta}{s}=\frac{2+\frac{\sum_{i}{q_{i}}}{r_{h}^{2}}
-\frac{\prod_{i}{q_{i}}}{r_{h}^{6}}}{8\pi\prod_{i}(1+\frac{q_{i}}{r_{h}^{2}})^{\frac{1}{3}}}
 \left[1-{\frac{\sum_{i} q_{i}}{{3r_{{h}}}^{2}}} + {\frac {5\sum_{i} q_{i}^{2}+ 4\sum_{i\neq j}
 q_{{i}}q_{{j}}}{{27r_{{h}}}^{4}}}
- \cdots \right],
\end{equation}
where $r_{h}$ is root of the equation $f(r)=0$ (and we set $\frac{1}{\lambda^{2}}=2$). Here, the first term in the relation (8) for $q_{i}=0$, agree with
universality of the ratio of shear viscosity to entropy which is $\eta/s=1/4\pi$. Our main goal is computing the ratio of shear viscosity to entropy from
STU model for three different charges. Now we are going to discuss several aspect of this theory, so we consider three special cases. In the first step we
assume $q_{1}=q, q_{2}=q_{3}=0$. In that case from the relation (8) we have,
\begin{equation}\label{9}
\frac{\eta}{s}=\frac{2+\frac{q}{r_{h}^{2}}}{8\pi(1+\frac{q}{r_{h}^{2}})^{\frac{1}{3}}}
\left[1+\mathcal{O}(q)
 \right],
\end{equation}
where horizon radius obtained as the following (root of the
$f(r)=0$),
\begin{equation}\label{10}
r_{h}=\frac{1}{2}\,\sqrt {-2\,q+2\,\sqrt {{q}^{2}+8\ \mu}},
\end{equation}
By inserting the equation (10) into the relation (9) we can obtain $\eta/s$ in terms of black hole charge. So we draw figure of the ratio $\eta/s$ in terms
of $q$ for three different space in the Fig. 1. It shows us that $\frac{\eta}{s}$ is acceptable for all of values of $q$. The left plot of the Fig. 1 has
been drawn for the small black hole charge. Thermodynamical stability of this case reduces to $\frac{q}{r_{h}^{2}}<2$ [32]. Extracting $r_{h}$ in terms of
$T$ and $q$ from the relation (4) lead us to choose the black hole charge of order $10^6$ for $T=300 MeV$. It seems that our result is in contrast with the
previous studies such as the Ref. [10], where $\eta/s=1/4\pi$ verified. Therefore one conclude wrongly that our starting point, the relation (6), is not
applicable for the recent background. However, we would like to write special condition where our results coincide with the Ref. [10]. If we set
$q/r_{h}^{2}=18/29$, then $\eta/s=1/4\pi$ and this ratio is independent of $q$. We should note that the above condition agree
with the thermodynamical stability.\\

\begin{figure}[th]
\begin{center}
\includegraphics[scale=.35]{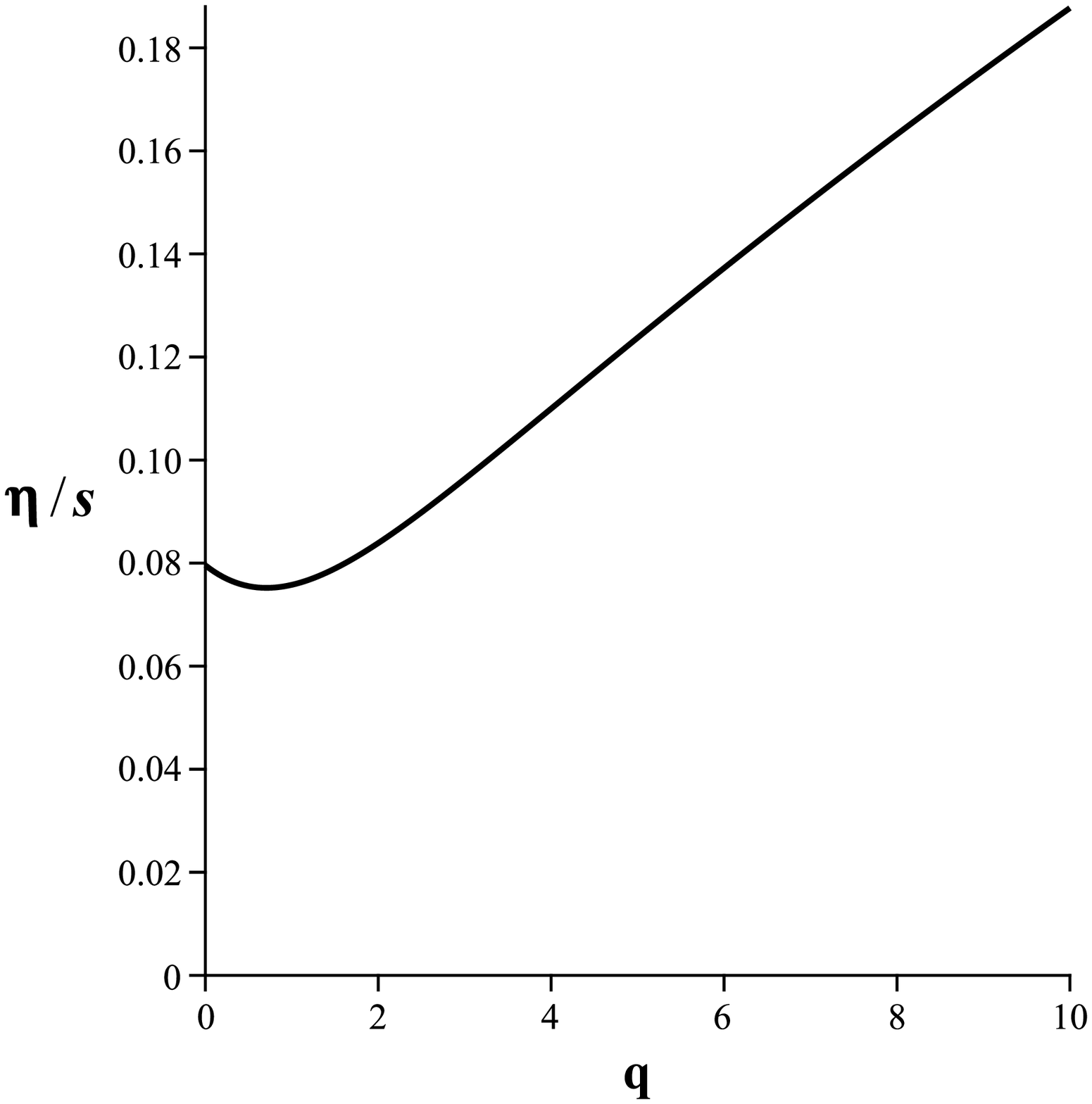}\includegraphics[scale=.35]{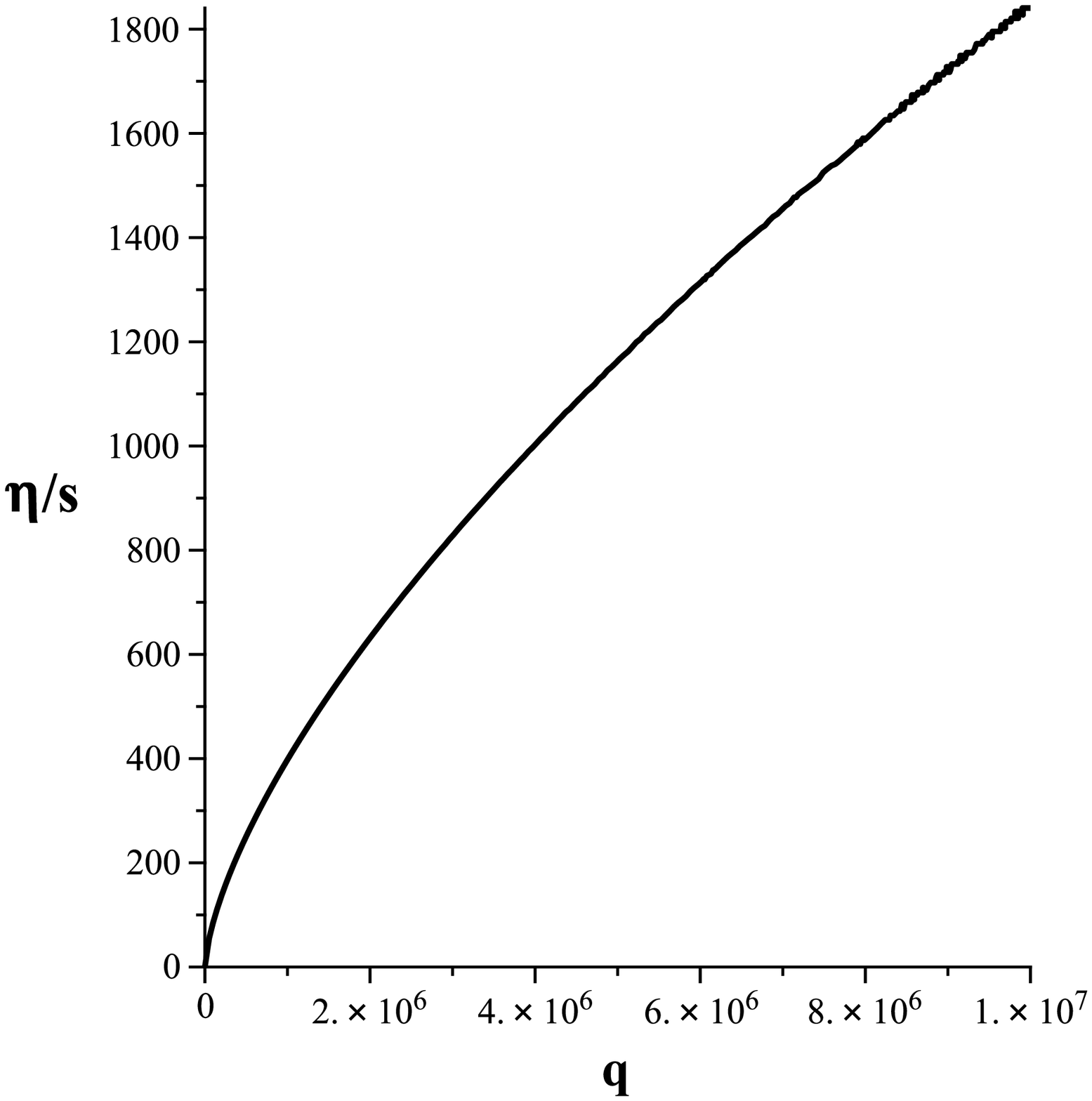}
\caption{The graphs of $\eta /s$ for the case of $q_{1}=q, q_{2}=q_{3}=0$ by choosing $\lambda=\sqrt{2}/2$ and $\mu=0.5$. Left: For the small black hole
charge. Right: For the large black hole charge.}
\end{center}
\end{figure}

\begin{figure}[th]
\begin{center}
\includegraphics[scale=.35]{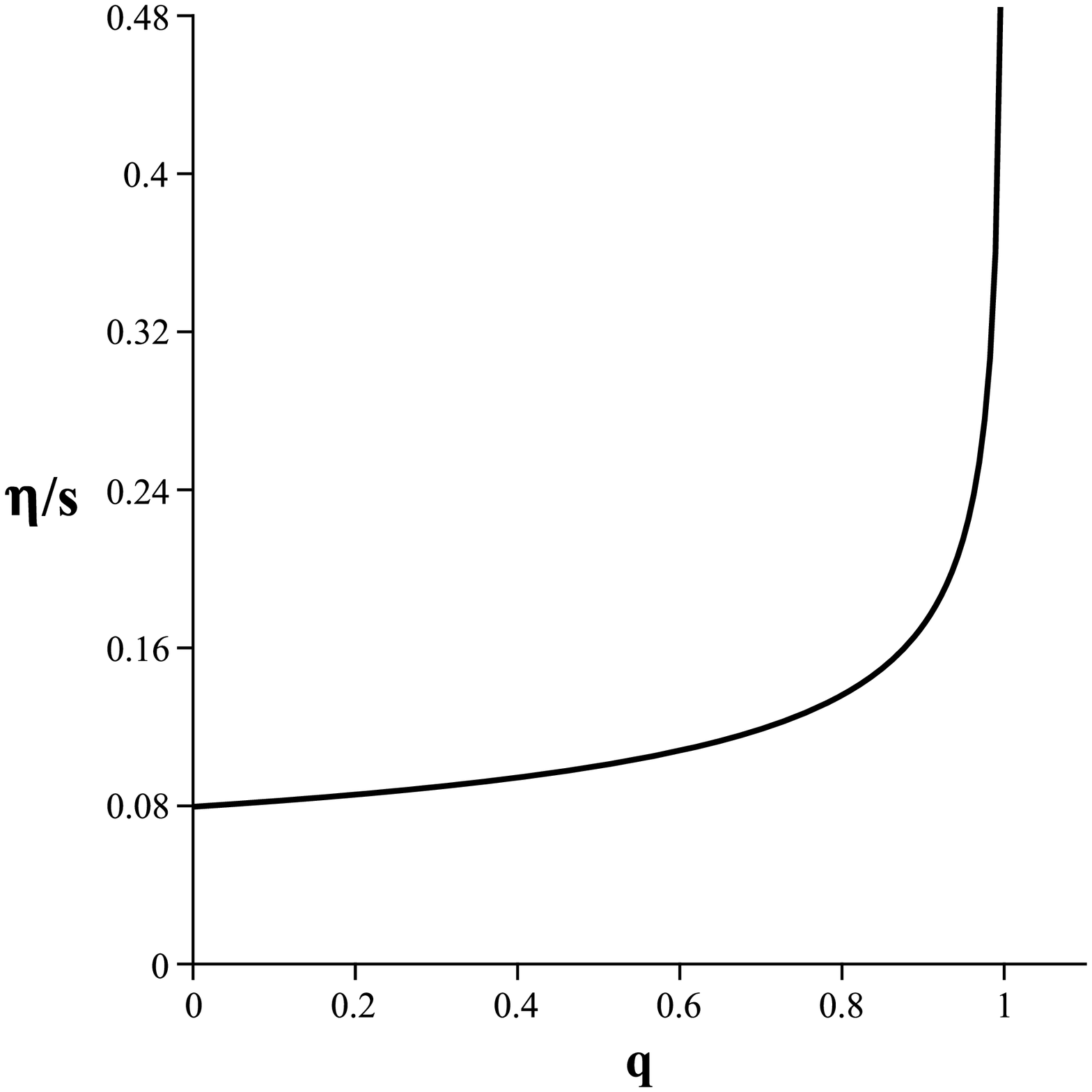}
\caption{The graphs of $\eta /s$ for the case of $q_{1}=q_{2}=q, q_{3}=0$ by choosing $\lambda=\sqrt{2}/2$ and $\mu=0.5$.}
\end{center}
\end{figure}

In the second step we consider $q_{1}=q_{2}=q, q_{3}=0$. In that case from the relation (8) we have,
\begin{eqnarray}\label{11}
\frac{\eta}{s}=(1+\frac{q}{r_{h}^{2}})^{\frac{1}{3}} \frac{1}{4\pi}
\Big[1+\mathcal{O}(q)
 \Big],
\end{eqnarray}
where,
\begin{equation}\label{12}
r_{h}=\sqrt {-q+\sqrt {2\,\mu}}.
\end{equation}
By inserting  the equation (12) into the relation (11) we can obtain $\eta /s$ in terms of black hole charge, so we plot the graph of $\eta /s$ in terms of
$q$ in the Fig. 2. It shows us that $\eta /s$ is acceptable for $q\leq 1$. Thermodynamical stability of this case reduces to $\frac{q}{r_{h}^{2}}<1$.
Extracting $r_{h}$ in terms of $T$ and $q$ from the relation (4) lead us to choose the black hole charge of order $10^6$ for $T=300 MeV$. It means that the
thermodynamical stability of two charged black hole satisfies with less charge than one charge black hole. For the large value of the black hole charge
(larger than one) the ratio of the shear viscosity to entropy goes to infinity. In that case by choosing $q/r_{h}^{2}=9/26$ we get $\eta/s=1/4\pi$ and
there is no dependence on $q$ so we recover results of the Ref. [10]. We should
note that this value of $q/r_{h}^{2}$ agree with the thermodynamical stability ($q/r_{h}^{2}<1$).\\

\begin{figure}[th]
\begin{center}
\includegraphics[scale=.35]{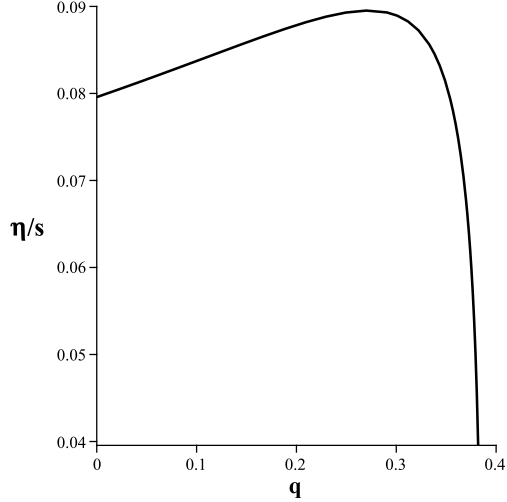}
\caption{The graphs of $\eta /s$ for the case of $q_{1}=q_{2}=q_{3}=q$ by choosing $\lambda=\sqrt{2}/2$ and $\mu=0.5$.}
\end{center}
\end{figure}

In the third step we assume that $q_{1}=q_{2}=q_{3}=q$. In that case
from the relation (8) we have,
\begin{eqnarray}\label{13}
\frac{\eta}{s}=\frac{2+\frac{3\,{q}}{r^2_h}-\frac{{q^{3}}}{r^6_h}}{8\pi(1+\frac{q}{r_h^2})}
\Big[1+\mathcal{O}(q)
 \Big],
\end{eqnarray}
where,
\begin{equation}\label{14}
r_{h}=\frac{\sqrt{3}}{3}\,\sqrt{\mathcal{M}^{\frac{1}{3}}-6\,\mu\,\mathcal{M}^{\frac{-1}{3}}-3\,q},
\end{equation}
where,
\begin{eqnarray}\label{15}
\mathcal{M}=-27\,\mu\,q+3\,\sqrt
{-24\,{\mu}^{3}+81\,{\mu}^{2}{q}^{2}},
\end{eqnarray}
By inserting  the relation (15) into the equation (13) we can obtain $\eta /s$ in terms of black hole charge, so we plot the graph of $\eta /s$ in terms of
$q$ in the Fig. 3. It shows that $\eta /s$ is acceptable for $q$ in the range of $0<q<0.35$. Thermodynamical stability of this case reduces to
$\frac{q}{r_{h}^{2}}(3-\frac{q^2}{r_{h}^{4}})<2$. Extracting $r_{h}$ in terms of $T$ and $q$ from the relation (4) lead us to choose the black hole charge
of order $10^6$ for $T=300 MeV$, but the large value of the black hole charge yields to the negative $\eta /s$. This result agree with the result of the
Ref. [21] where turning on R-charge leads to violation of the $\eta /s$ bound. In the next section we see that the negative part of $\eta /s$ can be left
by adding the higher derivative terms. If we choose $q/r_{h}^{2}=1/33$ then we find $\eta/s=1/4\pi$ and we are agree with the results of the Ref. [10].
\section{Higher derivative correction}
Now, we ready to calculate the effect of higher derivative corrections. Higher derivatives of STU model already studied (for the case of $k=1$) [20], and
applied to $\eta/s$ for the black hole with three equal charge ($q_{1}=q_{2}=q_{3}=q$) [21]. Now we would like to extend this work to the case of three
different charges in the flat space. In that case the metric (1) reminds unchange but,
\begin{equation}\label{16}
H_i=1+\frac{q_i}{r^2}-\frac{cq_{i}(q_{i}+\mu)}{72r^{2}(r^{2}+q_{i})^{2}},
\end{equation}
and
\begin{eqnarray}\label{17}
f&=&-\frac{\mu}{r^2}+\lambda^2 r^2
\prod_{i}(1+\frac{q_i}{r^2})\nonumber\\
&+&c\left(\frac{\mu^{2}}{96r^{6}\prod_{i}(1+\frac{q_i}{r^2})}
-\frac{\lambda^{2}\prod_{i}q_{i}(q_{i}+\mu)}{9\,r^{4}}\right),
\end{eqnarray}
where $c$ is small constant parameter corresponding to higher
derivative terms. In that case the modified horizon radius is given
by,
\begin{eqnarray}\label{18}
r_{h}&=&r_{0h}\nonumber\\
&+&\frac{c}{24}\frac{\lambda^{4}(\prod_{i}(1+\frac{q_i}{r_{0h}^2}))^{\frac{4}{3}}(\sum
q_{i}^{2}-\frac{26 r_{0h}^{2}}{3}\sum
q_{i}+3r_{0h}^{4})}{24{(\prod_{i}(1+\frac{q_i}{r_{0h}^2}))}^{\frac{1}{3}}
\left[\lambda^{2}(\prod_{i}(1+\frac{q_i}{r_{0h}^2}))^{\frac{2}{3}}(\frac{1}{3}\sum
q_{i}-2 r_{0h}^{2})-1\right]}\nonumber\\
&-&\frac{c}{24}\frac{2\lambda^{2}(\prod_{i}(1+\frac{q_i}{r_{0h}^2}))^{\frac{2}{3}}(\frac{13}{3}\sum
q_{i}-3
r_{0h}^{2})+3}{{24{(\prod_{i}(1+\frac{q_i}{r_{0h}^2}))}^{\frac{1}{3}}
\left[\lambda^{2}(\prod_{i}(1+\frac{q_i}{r_{0h}^2}))^{\frac{2}{3}}(\frac{1}{3}\sum
q_{i}-2 r_{0h}^{2})-1\right]}},
\end{eqnarray}
where $r_{0h}$ is the horizon radius without higher derivative
corrections.\\
In that case the black hole temperature is given by the following
relation,
\begin{equation}\label{19}
T=\frac{1}{4\pi} f'(r_h) \Big(H_1 H_2 H_3\Big)^{-\frac{1}{2}}
\end{equation}
where prime denotes derivative with respect to $r$, and $f(r)$ is
given by the relation (17). So, we obtain diffusion constant as the
following form,
\begin{equation}\label{20}
D=\prod^3_{i=1}(1+\frac{q_i}{r_h^2}-\frac{cq_{i}(q_{i}+\mu)}{72r^{2}(r^{2}+q_{i})^{2}})^{\frac{1}{6}}
\Big[\frac{1}{2\, r_h} - {\frac {\sum q_i}{{6\, r_{{h}}}^{3}}} +
\mathcal{O}(\frac{1}{r_{h}^{5}})\Big].
\end{equation}
Hence we get,
\begin{equation}\label{21}
\frac{\eta}{s}=\frac{f'(r_h)}{{4\pi\prod_{i}(1+\frac{q_i}{r_h^2}-\frac{cq_{i}(q_{i}+\mu)}
{72r^{2}(r^{2}+q_{i})^{2}})^{\frac{1}{3}}}}\frac{1}{2\,
r_h}\Big[1 - {\frac {\sum q_i}{{3\, r_{{h}}}^{2}}}
+\mathcal{O}(\frac{1}{r_{h}^{4}}) \Big].
\end{equation}
We plot the $\eta / s$ in terms of $q$ in the Fig. 4. Three different figures show that the higher derivative terms increase the
ratio of $\eta / s$, but the leading order ratio $\eta / s=1/4\pi$ is universal.\\
\begin{figure}[th]
\begin{center}
\includegraphics[scale=.23]{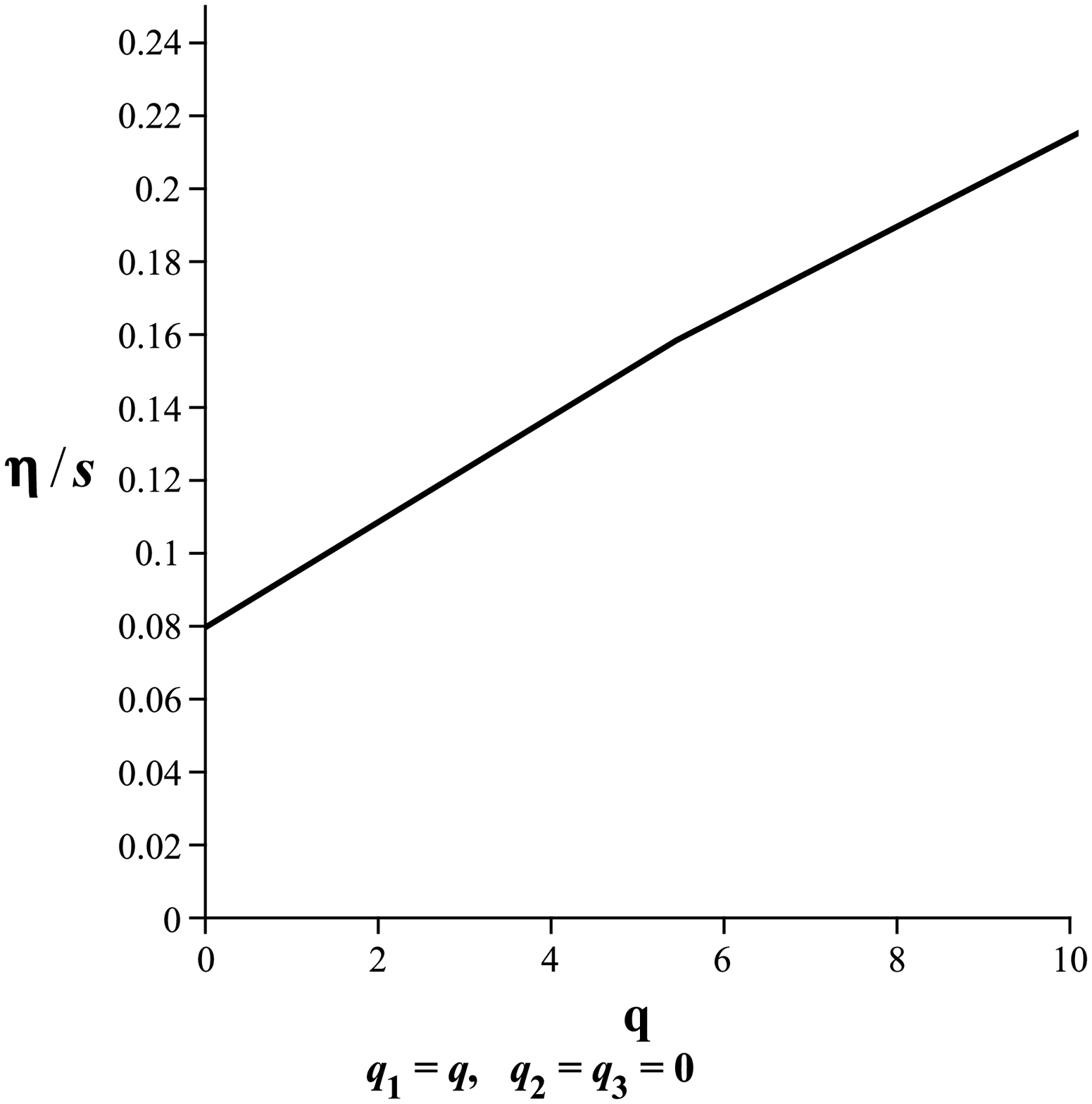}\includegraphics[scale=.23]{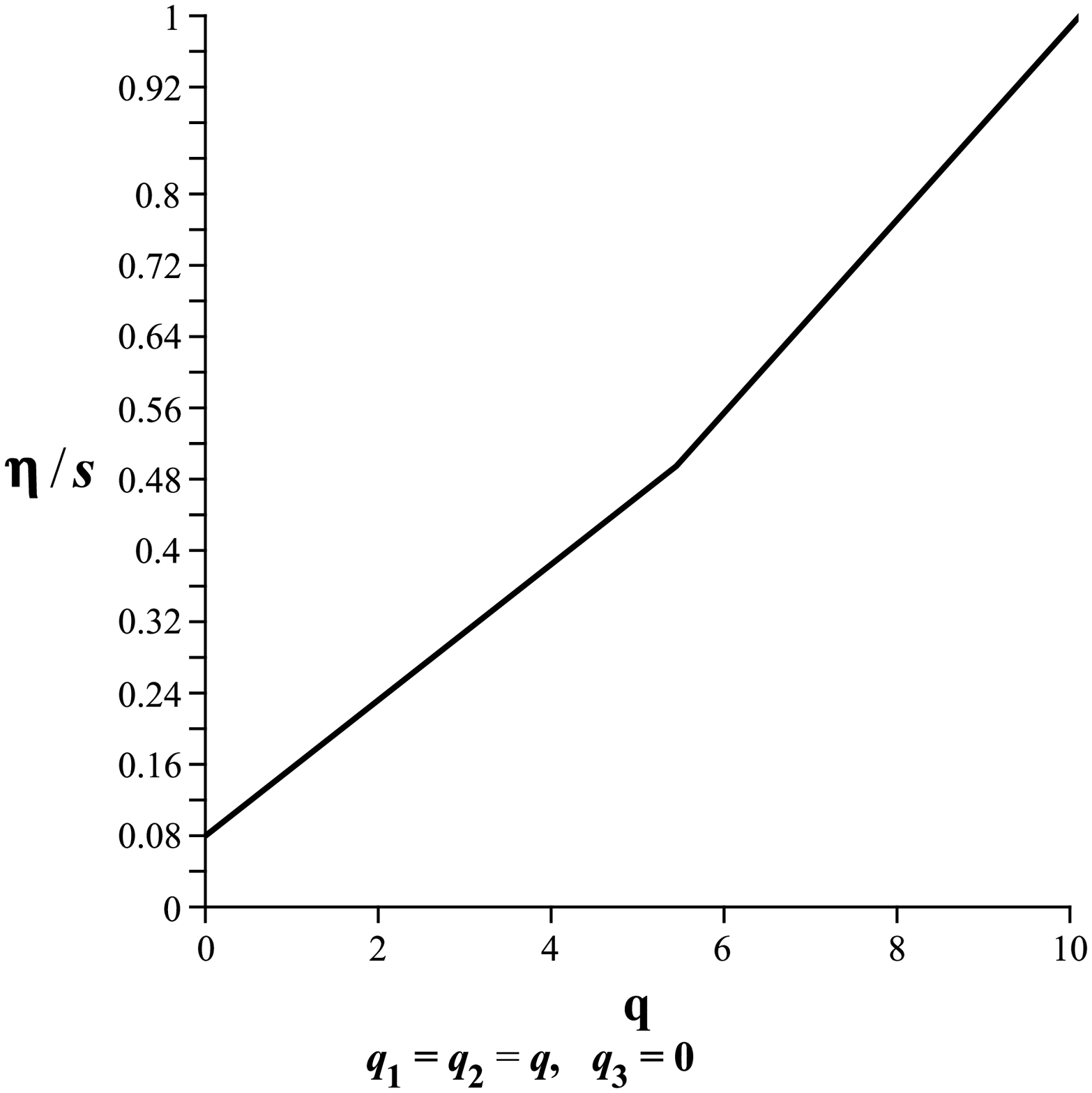}
\includegraphics[scale=.23]{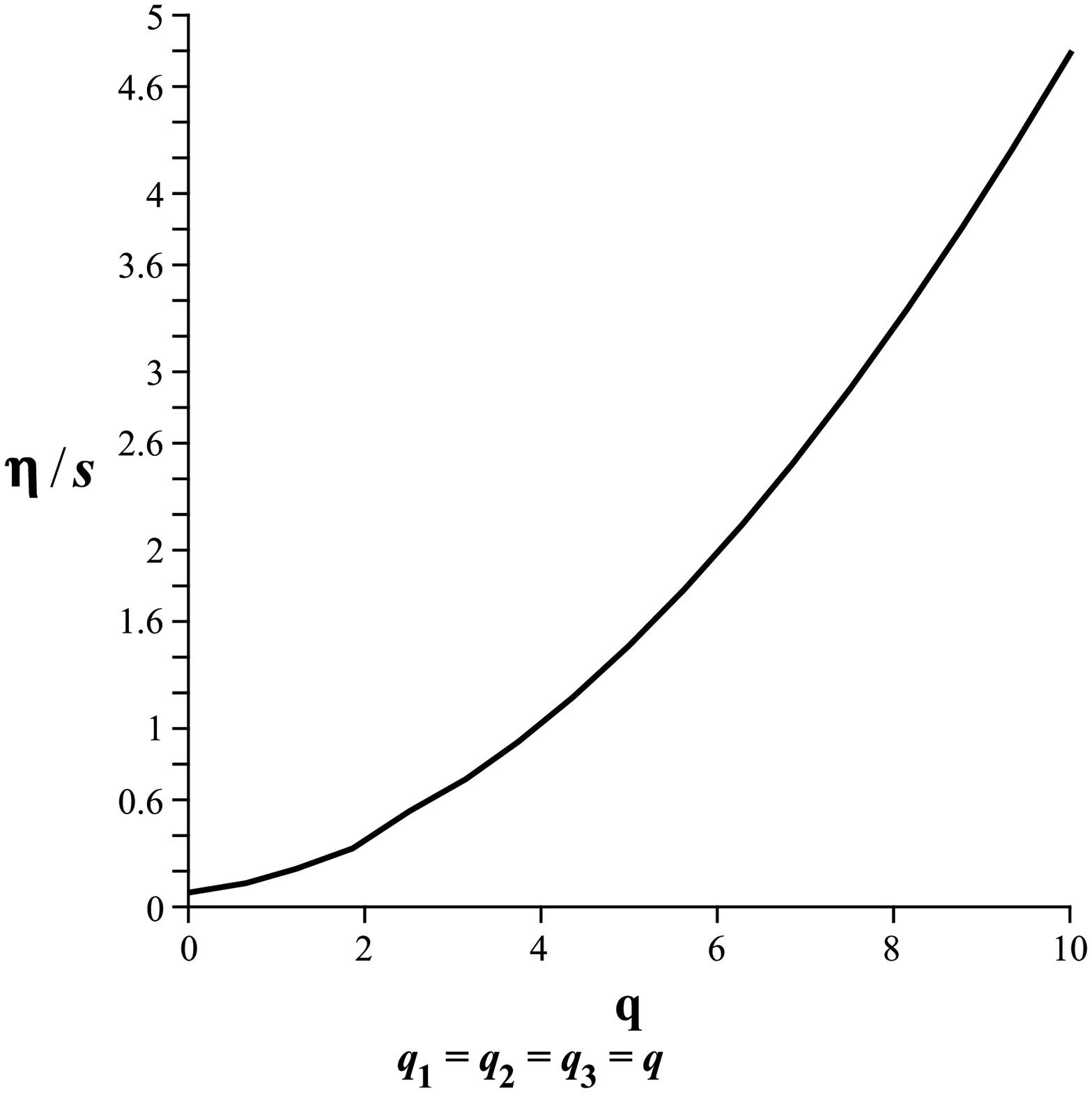}
\caption{The graphs of $\eta /s$ for three different cases by
choosing $\lambda=\sqrt{2}/2$, $\mu=0.5$, and $c=0.05$.}
\end{center}
\end{figure}

\section{Conductivity}
Now we would like to use results of the [36] to obtain thermal and electrical conductivities. In the Ref. [36] it is found that the conductivities for
gauge theories dual to R-charge black hole in 4, 5 and 7 dimensions behaves in a universal manner.

\begin{figure}[th]
\begin{center}
\includegraphics[scale=.3]{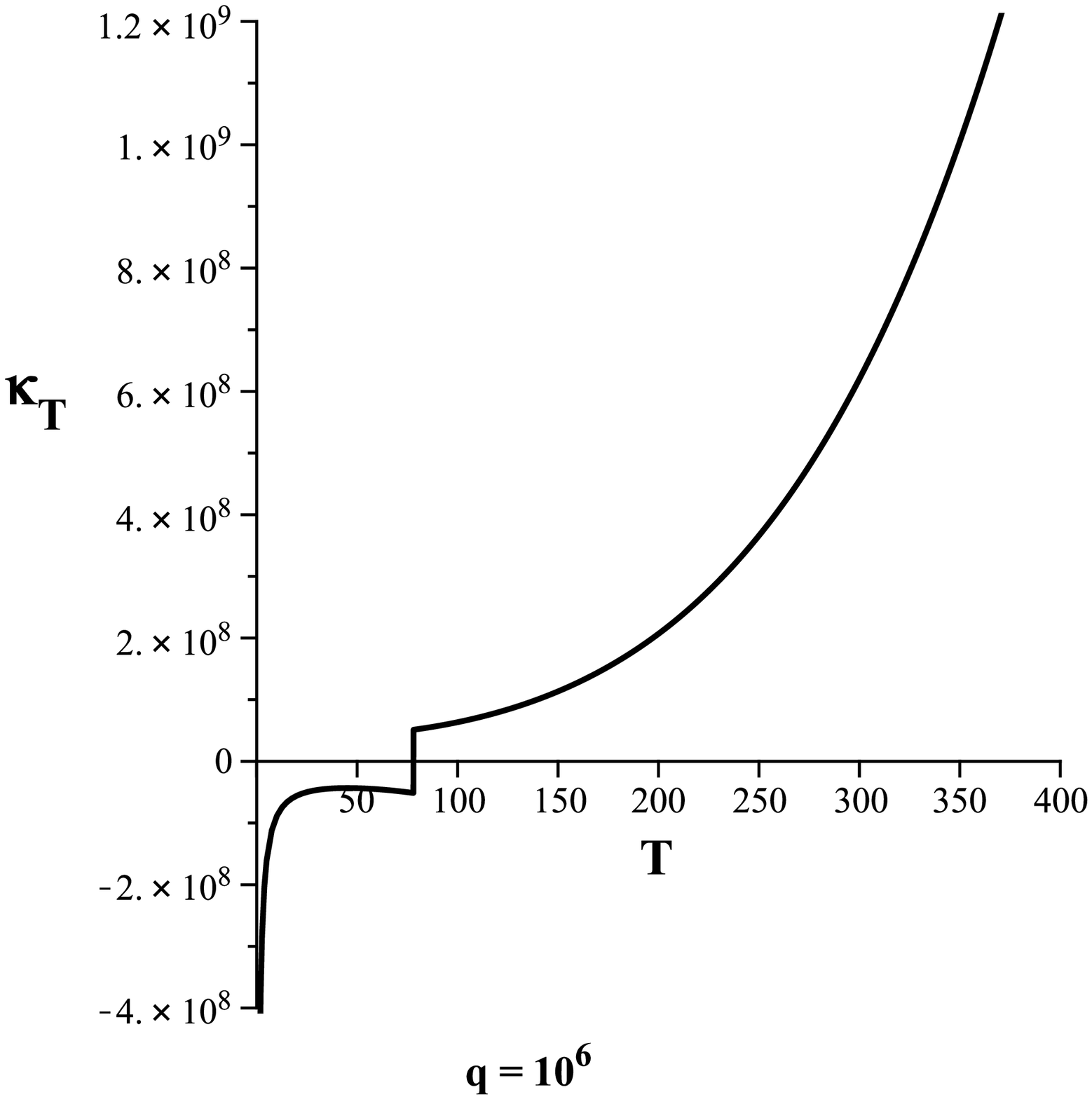}\includegraphics[scale=.3]{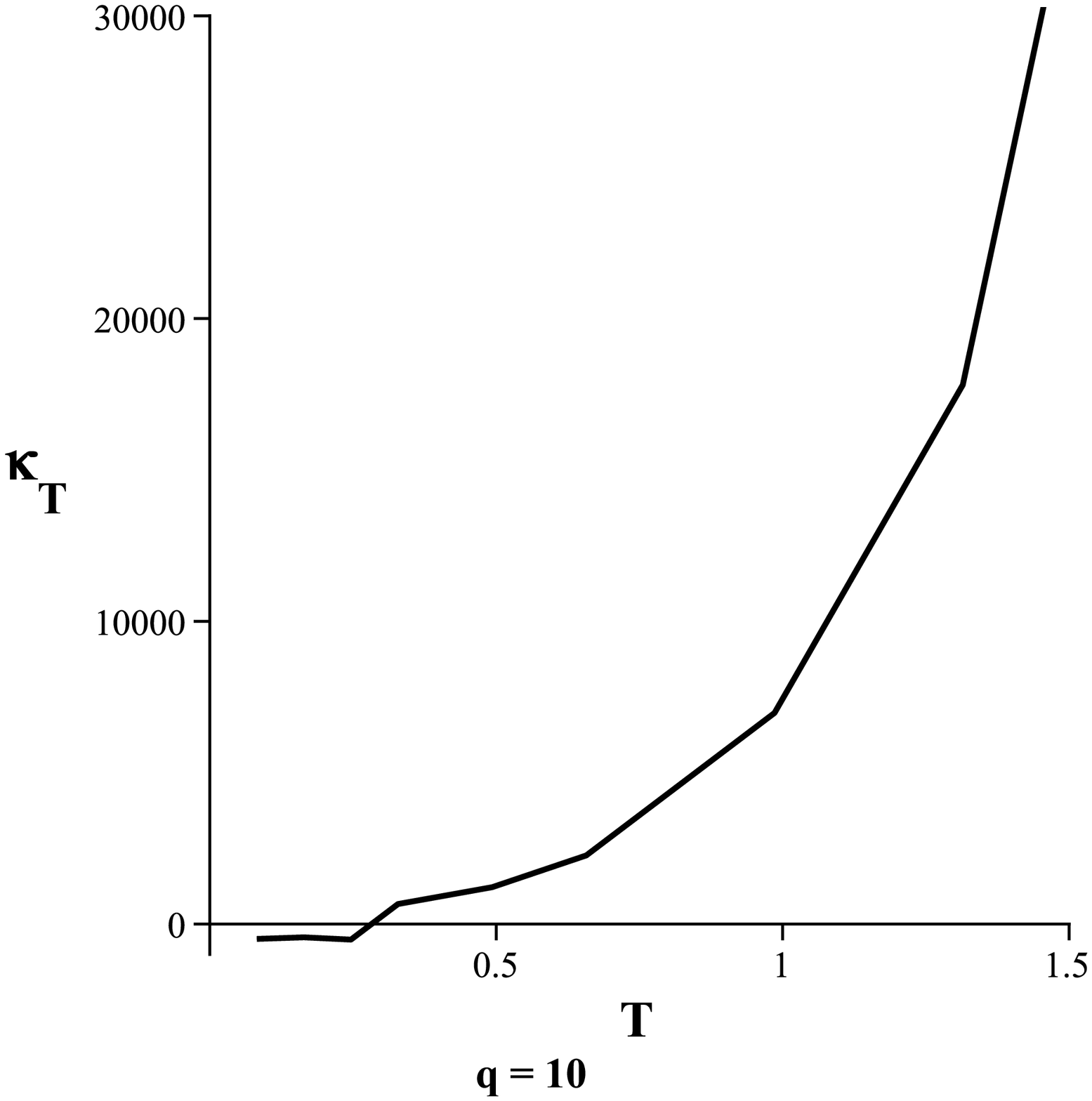}
\caption{The graphs of $\kappa_{T}$ for the case of $q_{1}=q,
q_{2}=q_{3}=0$. We set the parameters as $q=10^6$ and $\mu=0.5$.}
\end{center}
\end{figure}

According to the Ref. [36] and using line element (1) one can obtain,
\begin{equation}\label{22}
\sigma_{H}=\frac{r_{h}}{\lambda^{2}}\prod_{i}\left(1+\frac{q_{i}}{r_{h}^{2}}\right)^{\frac{3}{2}}.
\end{equation}
Also thermal conductivity obtained as the following expression,
\begin{equation}\label{23}
\kappa_{T}=\left(\frac{\epsilon+P}{\rho}\right)^{2}\frac{\sigma_{H}}{T},
\end{equation}
where the energy density $\epsilon$, pressure $P$ and density of
physical charge are defined as [41],
\begin{eqnarray}
\epsilon&=&\frac{3N^{2}r_{h}^{4}\lambda^{8}}{8\pi^{2}}\prod_{i}\left(1+\frac{q_{i}}{r_{h}^{2}}\right),\label{24}  \\
P&=&\frac{N^{2}r_{h}^{4}\lambda^{8}}{8\pi^{2}}\prod_{i}\left(1+\frac{q_{i}}{r_{h}^{2}}\right),\label{241} \\
\rho&=&\frac{\sqrt{2\sum_{i}q_{i}}N^{2}r_{h}^{2}\lambda^{6}}{8\pi^{2}}\sqrt{\prod_{i}\left(1+\frac{q_{i}}{r_{h}^{2}}\right)},
\label{242}
\end{eqnarray}
where $N^{2}=\frac{8\pi^{2}}{\lambda^{3}}$, and we used $8\pi G=1$. In the Fig. 5 we draw graph of $\kappa_{T}$ in terms of the temperature for the
simplest case of $q_{1}=q, q_{2}=q_{3}=0$. It shows that the thermal conductivity vanishes at $T\approx28 MeV$ for large black hole charge and $T\approx0.9
MeV$ for small black hole charge. It means that the thermal conductivity decrease with the black hole charge.

\begin{figure}[th]
\begin{center}
\includegraphics[scale=.3]{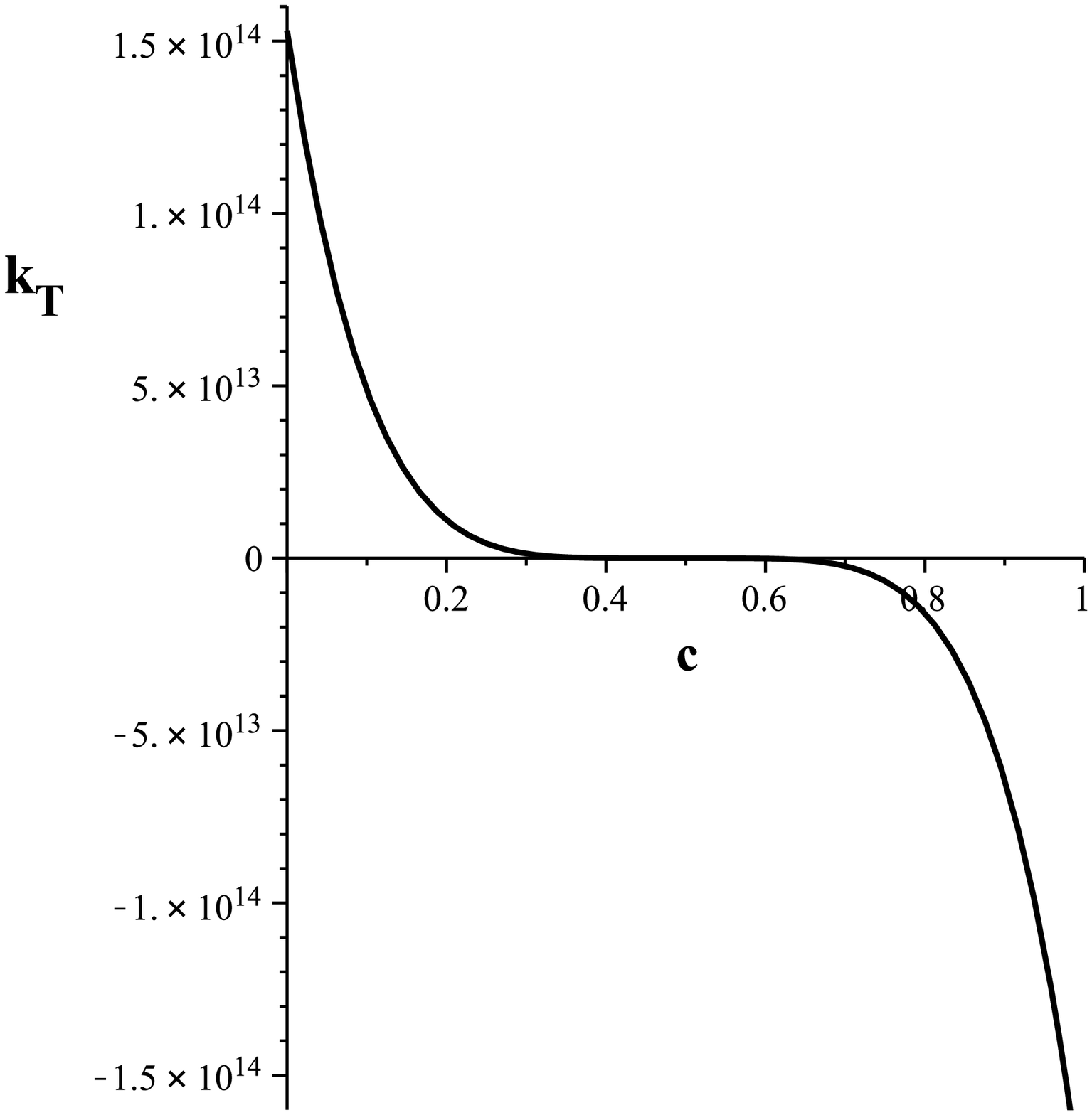}\includegraphics[scale=.3]{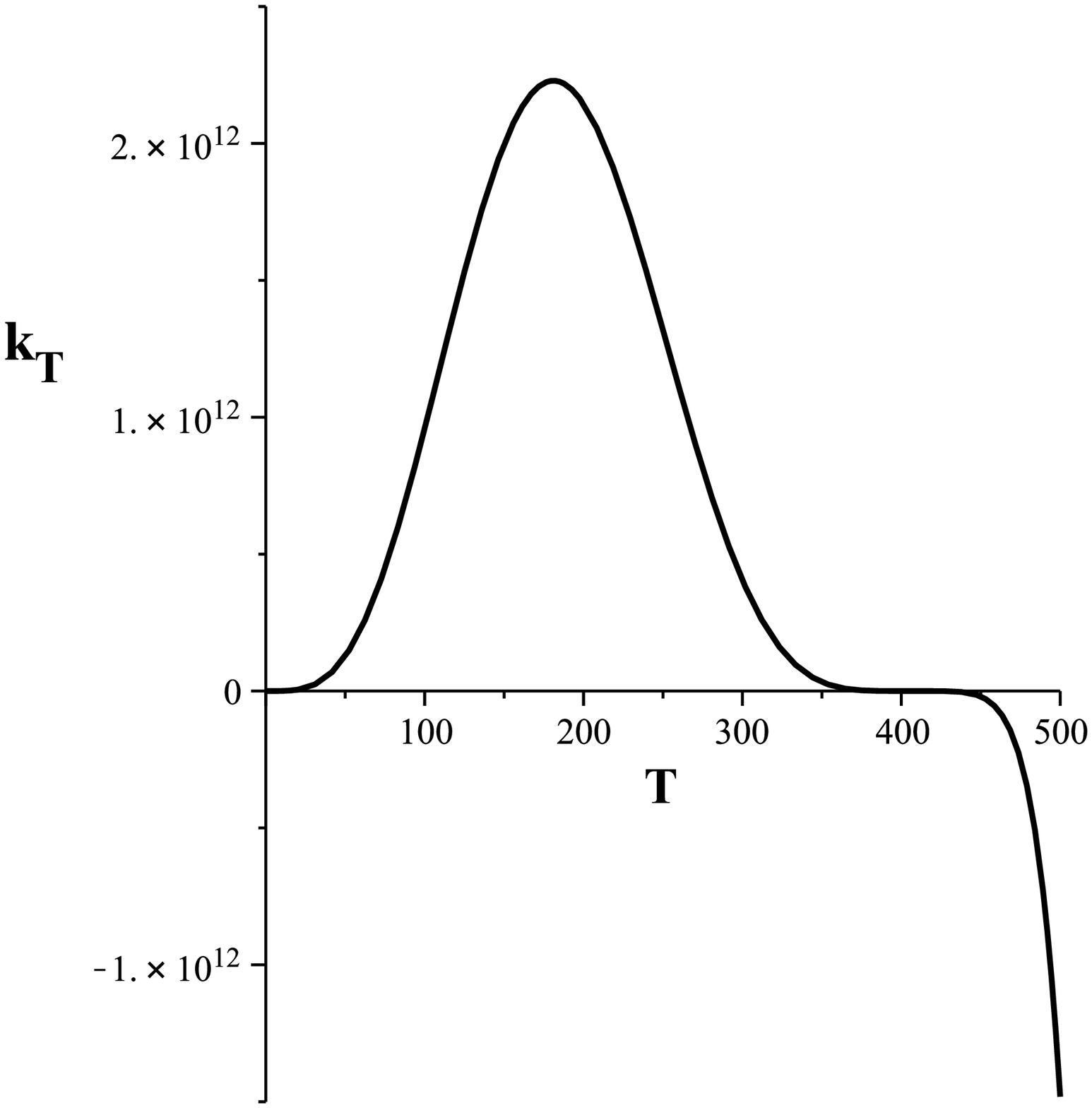}
\caption{The graphs of $k_T$ for the higher derivative correction by choosing $\lambda=\sqrt{2}/2$, $\mu=0.5$. Left: Thermal conductivity in terms of $c$
for $T=250 MeV$. Right: Thermal conductivity in terms of $T$ for $c=0.3$}
\end{center}
\end{figure}

The extension of the relations (22) and (23) to include the higher derivative terms is more complicated. Therefore just we draw graph of the thermal
conductivity in terms of the higher derivative parameter $c$ (the left plot of the Fig. 6), and in terms of the temperature (the right plot of the Fig. 6).
These figures tell us that the large value of the higher derivative parameter yields to the negative thermal conductivity, which is not acceptable. For
example in the case of $T=250 MeV$ one can obtain $c\leq0.6$. In that case for the case of $c=0.3$ the thermal conductivity becomes negative for $T>450
MeV$.
\section{Shear viscosity from the density of physical charge}
In this section, we are going to obtain shear viscosity as the
following form,
\begin{equation}\label{eta2s}
    \frac{\eta}{D}=s\,T+\mu \, \rho
\end{equation}
where
\begin{equation}\label{s}
    s=\frac{N^2\,\lambda^6}{2\,\pi}r_h^3\,\prod_{i=1}^3\left(1+\frac{q_i}{r_h^2}\right)^{\frac{1}{2}}.
\end{equation}
By inserting Eqs. (4), (7) and (28) into the Eq. (27), the $\eta / s$ becomes,
\begin{eqnarray}\label{eta2s1}
    &\frac{\eta}{s}=\frac{1}{4\,\pi}\left[\frac{2+\frac{q_1+q_2+q_3}{r^2_h}-\frac{q_1\,q_2\,q_3}{r^6_h}
    }{\prod_{i=1}^3\left(1+\frac{q_i}{r_h^2}\right)^{\frac{1}{2}}
    }+\frac{\sqrt{2}\,\mu\,\lambda^4\,\sqrt{q_1+q_2+q_3}}{2\,\pi^2}\right]\prod_{i=1}^3\left(1+\frac{q_i}{r_h^2}
    \right)^{\frac{1}{6}},
\end{eqnarray}
now we insert different $r_h$ from Eqs. (10), (12) and (14) into the Eq. (29) and draw $\eta / s$ in terms of $q$ in Figs. 7. The obtained consequence from
Fig. 7 shows that one equals with previous consequence.

\begin{figure}[th]\label{etasec}
\begin{center}
\includegraphics[scale=.23]{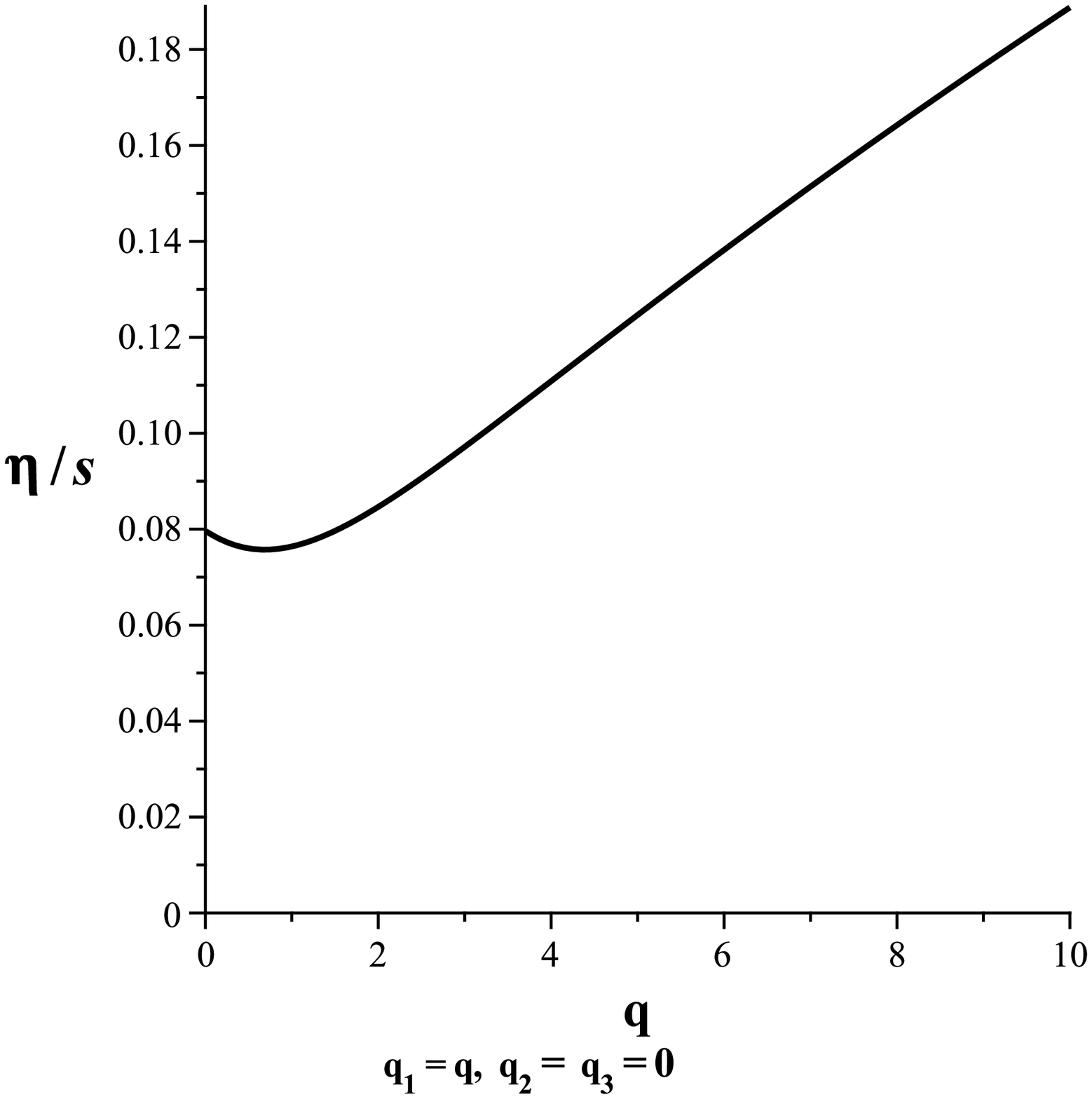}\includegraphics[scale=.23]{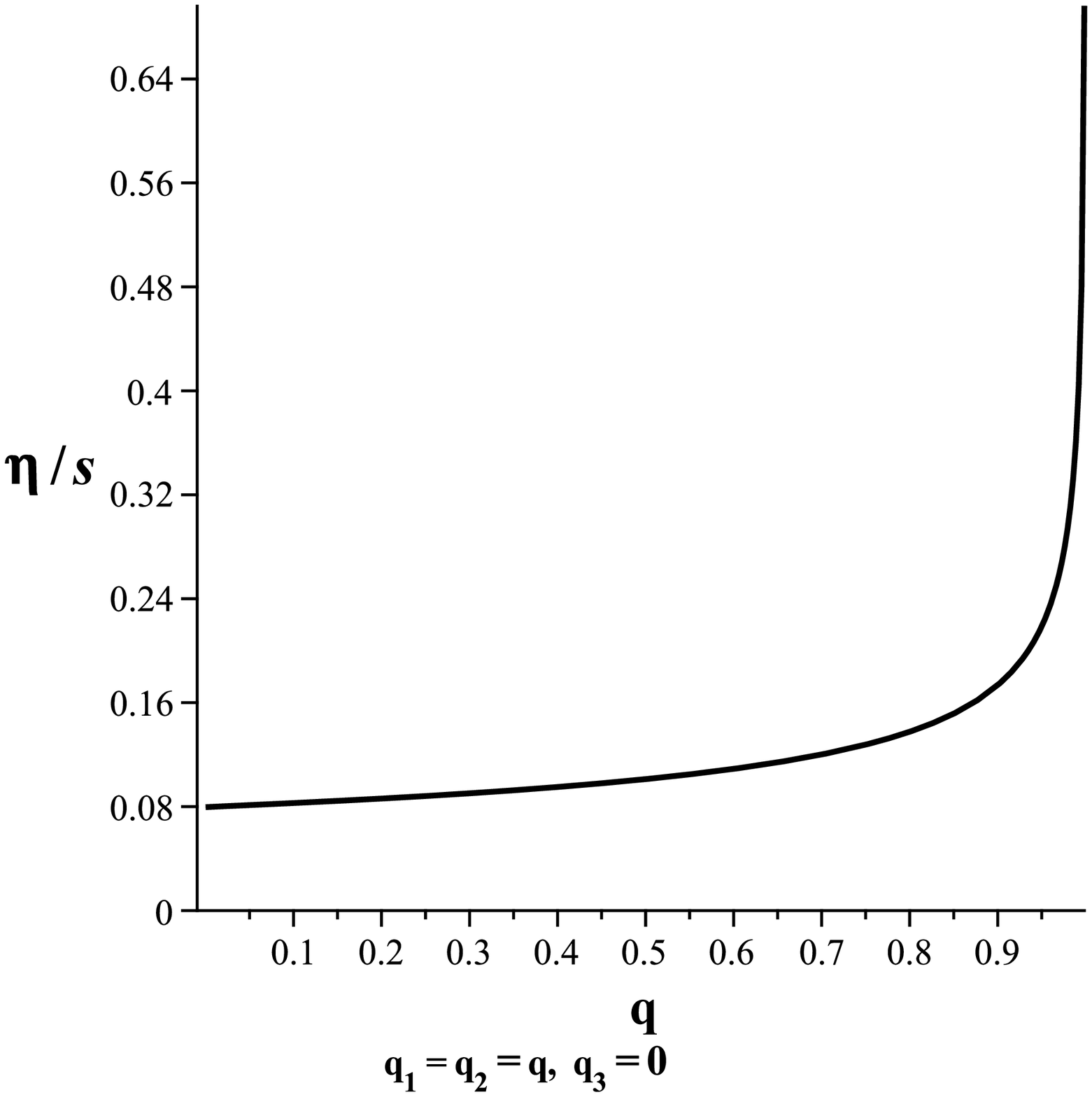}
\includegraphics[scale=.23]{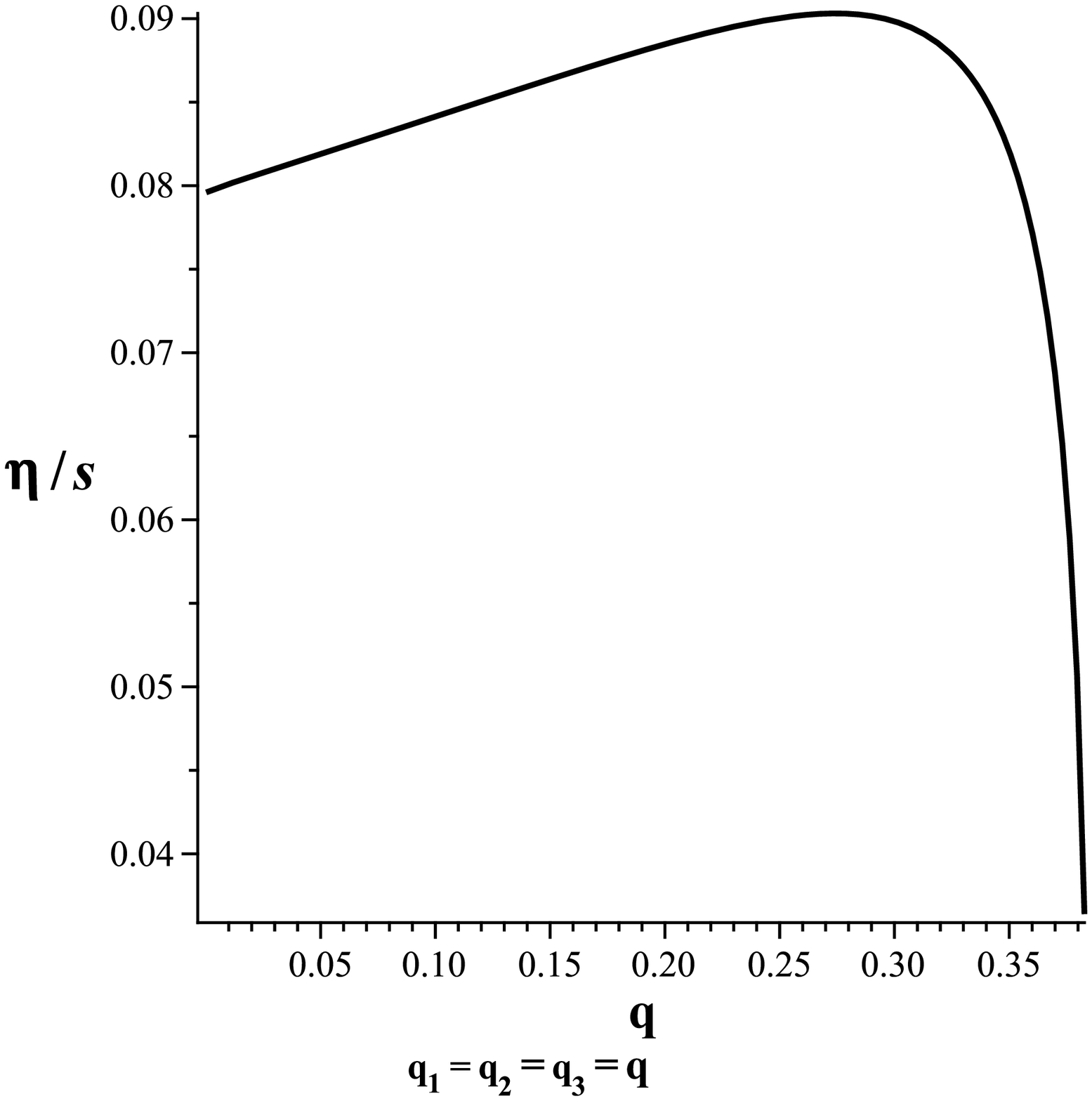}
\caption{The graphs of $\eta/s$ for three cases black hole charge configuration.}
\end{center}
\end{figure}

\section{Conclusion}
In this paper we considered the three-charged non-extremal black hole solution in a $\mathcal{N} = 2$ gauged supergravity (STU model) with arbitrary black
hole charges, and investigated some important hydrodynamical properties. By using the diffusion constant we obtained an analytical expression for the ratio
of shear viscosity to entropy density in the relation (8), then by using the thermodynamical stability discussed the special values of the black hole
charges. In all cases, to leading order, the ratio $\eta / s$ is universal. For the higher order terms include black hole charge we found special
constraints on the black hole charge which yield to universal value of $\eta/s$. We found for the black hole with three equal charges the lower bound of
$\eta / s$ violated for arbitrary value of the black hole charges. However the condition $\eta / s\geq1/4\pi$ satisfies if $0<q<0.25$. Mentioned violation
not happen for two and one-charged black holes. Also we calculated the thermal and electrical conductivity and found that the black hole charge decreases
the value of conductivity. Finally we considered the effect of the higher derivative correction and concluded that the ratio of the shear viscosity to
entropy grow with the higher derivative parameter. Therefore the higher derivative correction yield to validation of the conjectured $1/4\pi$ bound. This
situation is different for the thermal conductivity, the thermal conductivity decreases with the higher derivative parameter. So, for the large value of
the higher derivative parameter the thermal conductivity becomes negative. This yields us to obtain critical value of $c$ which depends on the temperature.
For example in the case of $T=250 MeV$ one can obtain $c\leq0.6$. By increasing the temperature the parameter $c$ take smaller value.

\end{document}